# Operational Aspects of Dealing with the Large BaBar Data Set

Tofigh Azemoon, Adil Hasan, Wilko Kröger, Artem Trunov
*SLAC Computing Services, Stanford, CA 94025, USA*
*On Behalf of the BaBar Computing Group*

To date, the BaBar experiment has stored over 0.7PB of data in an Objectivity/DB database. Approximately half this data-set comprises simulated data of which more than 70% has been produced at more than 20 collaborating institutes outside of SLAC. The operational aspects of managing such a large data set and providing access to the physicists in a timely manner is a challenging and complex problem. We describe the operational aspects of managing such a large distributed data-set as well as importing and exporting data from geographically spread BaBar collaborators. We also describe problems common to dealing with such large datasets.

## 1. INTRODUCTION

The BaBar data persistency was built using Objectivity/DB [1] - an object oriented database management system. Objectivity/DB stores data in federated databases, or federations, and provides many powerful DBMS features such as transaction management and distributed data access. The Objectivity/DB setup for BaBar is described in Section 2.

Early in the experiment it was quickly realized that the separation of production (predominantly writing) from analysis (predominantly reading) tasks was essential in order to maximize the performance of each task. To this end, analysis federations that are mirror copies of the production federations are created. The BaBar production system is described in Section 3. The analysis environment and operational issues, including data management are discussed in section Section 4. Section 5 deals with issues concerning the imports of databases from remote sites and Section 6 describes the tools used to perform data management at SLAC.

## 2. OBJECTIVITY/DB

The Objectivity/DB infrastructure consists of a federated database, or federation, including file catalog and database files (databases). The locking mechanism is provided by a lockserver, which is a separate application, running as a daemon. The same lockserver can be used by multiple federations. Standalone access to a federation without a lockserver is possible, but rarely used. Transaction journaling uses a dedicated server. Distributed data access is provided by the Advanced Multithreaded Server (AMS), running as a daemon on each host where data is placed. Distributed file access can be provided without the AMS, but all filesystems must be visible to the client, e.g. via NFS.

### 2.1. Using Objectivity/DB in BaBar

Objectivity/DB limits the number of databases in a federation to 64K. With our practical limit of 500 MB database file sizes, exhausting the database ids in a federations is an issue.

The Objectivity/DB catalog contains information for each of the database files that make up the federation. Each time a new database is created or introduced into a federation the catalog has to be updated. Since the beginning of the experiment it has been observed that catalog operations do not scale well with the increasing number of databases in a federation, making the 64K database id limit practically unreachable.

One of the essential requirements of database administration is to minimize downtime of the analysis federations. Since all access to a federation is through the catalog we have found it necessary to limit the maximum number of databases in a federation to 16K. In order to bypass the integrity checking that is already done in the production federations we attach databases to an analysis federation "by proxy", i.e. by attaching an empty database thereby bypassing the inherent Objectivity/DB integrity checking, and then replacing the empty file with the actual database file. The 16K limit coupled with the attach "by proxy" has helped us to achieve the goal of minimizing the downtime of an analysis federation. Slow catalog operations have been addressed by Objectivity/DB and should be greatly improved in the newer versions of Objectivity/DB starting with V8.0.

The lock server performance effectively limits the number of clients accessing a federation. The lock server is a single thread application and allows 1024 connections although fewer clients can also saturate the lockserver, depending on a client's activity. To reduce the load on a lockserver, Objectivity/DB introduced readonly databases, which greatly reduced the number of locks required by a client reading a database. The downside of this nice feature is that catalog operations on a readonly database require a quiescent federation.

SLAC runs a modified version of the AMS. Addi-tional development allowed multiple AMS processes to run on a single dataserver, increasing the overall data throughput. Also, the AMS is interfaced to a logical filesystem layer and a mass storage system, for better integration of all data access operations.

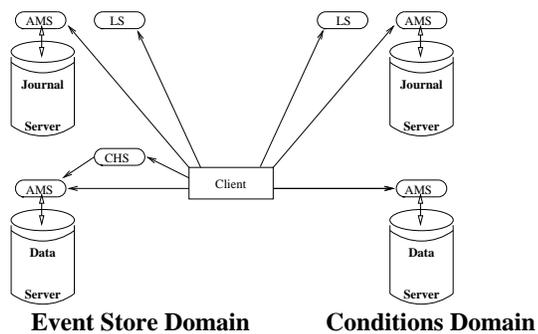

Figure 1: The prompt reconstruction setup.

A considerable improvement in the processing rate in data reconstruction was achieved by deploying a CORBA based clustering hint server (CHS) to pre-create a pool of empty pre-sized databases for a particular component. The server worked around two inherent problems in the database system: firstly, each time a new database needed to be created or a new container to be added, a write lock on the catalog, a central resource, was required. This caused other clients to stall waiting for the operation to be completed. The pre-creation of a pool of empty databases overcame this problem. Secondly, the time taken to read or write to the server disk increased over time. This was due to the server disk becoming fragmented as files were updated in small chunks. Pre-creating a full-sized empty database with the needed number of containers overcame this problem.

Access to the conditions information in data reconstruction was improved by introducing a CORBA based Object ID (OID) server, which cached object identifiers of navigational components of the conditions database, reducing access time and load on lockserver at start up time.

The production federation event store and conditions domains were placed in separate federations allowing each domain to have it's own lockserver. This had the effect of reducing the lockserver cpu utlilzation and allowed the system to handle a larger number of clients.

## 3. PRODUCTION ENVIRONMENT

### 3.1. Production Setup for Data Reconstruction

The production goal is to keep up with data coming from the BaBar detector. As luminosity increased, production had to scale accordingly.

In the initial design of the production system, use of only one production federation throughout the lifetime of the experiment was projected. This required file sizes to be 2 GB for micro level data and 20 GB for mini and raw data. About 60 clients processed data simultaneously and wrote into the production federation, which consisted of three data servers, a dedicated catalog host, journal and lockserver host.

With time, the production setup has changed to cope with the much increased luminosity and provide a more robust production environment. After introducing the OID and Clustering Hint Servers and moving the condition domain into a separate federation, the number of clients increased to 220, and the number of data server to six.

The current design of the data reconstruction system shown in Figure 1 takes advantage of two important changes in the BaBar offline software: the separation of the rolling calibrations from the data reconstruction and the reduction in the amount of data output for each event (only micro and mini-level information is written into a federation) [4].

The impact of the first change has been to allow for dedicated farms to be setup using enough hardware to process a sufficient sample of data in order to generate rolling calibrations needed by data reconstruction. There are currently three prompt calibration (PC) federations serving this year's data reconstruction and reprocessing of previous years' data at SLAC and Padova. Each PC federation resides on a dedicated data server, with it's own lockserver making every federation totally independent.

The second change allowed for data reconstruction federations to be concentrated on less hardware. Currently, each data reconstruction federation at SLAC uses one dedicated data server and one dedicated lockserver. The conditions needed to reconstruct the data are stored in a separate federation that resides on a different data server (see Figure 1). The conditions data server holds two conditions federations, each with it's own lockserver. The conditions federations used for the data reconstruction are readonly copies of the prompt calibration federations. There are currently five data reconstruction federations.

The new rolling calibration conditions needed for reconstruction are obtained by allowing the data reconstruction control system to copy the corresponding databases to the conditions federations. During the copy process the databases have to be quiescent, this is usually achieved by the reconstruction system pausing between runs to allow the copy to take place.

A similar environment has been setup at INFN-Padova for the reprocessing of BaBar data [5].

### 3.2. Production Setup for Simulation

In the simulation production environment, three collections were produced for each run, corresponding to the three stages of simulation production: simulation, background mixing and reconstruction. The simulation production infrastructure at SLAC includes three data servers accessed through the AMS, a dedicated catalog, lock and journal server.

About 100 concurrent jobs generate the data. The setup hasn't changed until recently, when the changes in the offline software (combining all stages of simulation in one application, removal of writing out intermediate data) allowed SLAC to introduce two more simulation production federations. Each of these federations reside on a single data server and each federation has its own lockserver. This system has been shown to handle over 200 clients without lockserver CPU saturation. These federations hold data produced by simulation applications, running in parasitic mode, on the same client machines as the data reconstruction applications. The simulation production capacity at SLAC should be further increased by adding 2 more federations to the setup.

The need for more simulation data was fulfilled by Monte Carlo production at remote sites. Remote sites usually have a different setup, driven by a site's limitation of resources. In the typical setup, a federation's metadata and generated data reside on single filesystem, which client applications accesses via NFS.

### 3.3. Data Management in Production

All data is stored in a High Performance Storage System (HPSS) developed by IBM. The AMS is interfaced to HPSS via the Objectivity Open File System (OOFS) - a cache file system developed at SLAC. The OOFS software allows automatic back up of new data, purging of old data from the file cache, and on-demand staging of archived data via the AMS.

In the production environment, data management is rather simple. Databases are forced to migrate to HPSS after being closed and are automatically purged off disk, in order to keep disk space usage low. Closed databases never need to be brought back to the production hosts. The production federation's metadata is backed up during production outages for data sweeps.

## 4. ANALYSIS ENVIRONMENT

### 4.1. Analysis Federations Setup

The analysis environment consists of over one hundred Objectivity/DB federations containing reconstruction or simulation data. Each analysis federation is a "mirror" of the corresponding production federation. To simplify data access for users, the concept of a "bridge" federation, combining families of federations was introduced. This provides a single point of entry to a family of federations [3]. For a user, a bridge federation looks like a normal analysis federation, but in reality it contains only pointers to collections in the "daughter" federations. Client applications switch to the correct daughter federation automatically. Bridge collections spanning across many daughter federations are also supported. Each analysis family uses conditions residing in a separate federation. Currently, four bridge federations are in use.

### 4.2. Data Transfer from Production Federations

One of an administrator's essential task is transferring the data from production to analysis federations, also known as "sweeping".

A sweep is achieved by copying database files from the production to the analysis servers. In order to ensure that the sweep is self consistent an outage of the production federation is required. An important requirement of the sweep is to have as little impact on the production federation as possible. In order to minimize the outage time the database files are closed to writing, allowing the weeks data production to be copied to the analysis federation whilst a new weeks production starts up writing to a new set of databases. The databases that need to be swept are found by comparing the production and analysis federation catalogs. The amount of data transferred during a sweep is between 300 to 500 GB.

Once the databases have been copied and registered in the analysis federation the collections need to be published allowing users to access the new data. The registering of new collections proceeds in two steps: registering of the collections in the daughter federation collection catalog followed by registering of the daughter collections in the bridge federation catalog. With an increased number of collections due to skimming being done as part of the data reconstruction, collection loading becomes a bottleneck in the sweep procedure as it takes a few hours to load the collections from a week's production.

Analysis federation users are not affected by an ongoing sweep and users learn about new data after it's been published in the metadata catalog. It is important to ensure the data quality before the collections are published for analysis. To this end QA applications are run against the new week's data identifying collections with corrupted data, missing databases etc.

### 4.3. Data Placement Strategy

In order to match the steady increase in the data volume, new disks are added to the servers used by the analysis federations (currently 50TB). It is desirable not to have "clusters" of old and new data, which affect the performance of analysis jobs. The data administration strategy here is to spread the data over many servers, potentially all data servers in the pool. Unfortunately, this increases the sensitivity of the analysis infrastructure to failure of a single data server. Relocation of data from one host to another also presents some difficulties: each database is registered in an Objectivity/DB catalog, and when a database is moved, the catalog has to be updated with it's new location. As a database is usually in a read-only state, such an update can only be made when the federation is quiescent.

### 4.4. Policies for Keeping Data on Disk

Until recently, the policy was to keep all micro level data on disk for immediate access. As data became old, it was pushed off disk. Separate data pools were allocated for micro level data which were permanently on disk, staged data, such as raw, which were kept on disk for a short period, and staged data, which were kept on disk for a long term. Staged data had to be prestaged by a user before a job could run.

Now, with an increased volume of data, it's not possible to hold the entire micro data set on disk. To combat this, a staging on demand mechanism was developed within the AMS. A client job stalls when the AMS server finds out that the requested database file is not on disk. At the same time, the AMS runs a staging program to transfer the file to disk. An independent purging daemon removes databases from disk based on their last access time. In Figure 2 typical access patterns, where the percentage of files last accessed (accumulated) versus last access time is shown. The histogram in the same figure shows the last access time of the files. Based on such a plot, a purging policy is set such that files not accessed for more than 10 days are purged when disk usage reaches 95%. Server pools for long and short term kept data were also merged and staging requests are managed dynamically.

### 4.5. Policies on User's Data

User's data includes data produced by individual users, by work groups, as well as official skims produced in the analysis environment from production collections. Since skim production puts significant stress on the analysis infrastructure, it's output is placed on dedicated servers, together with other user generated data. Official skim production and work group data is automatically backed up every week. During backup all jobs are stalled in order to guarantee that the federations data is quescent. Individual user's data is not systematically backed up in order not to overload HPSS.

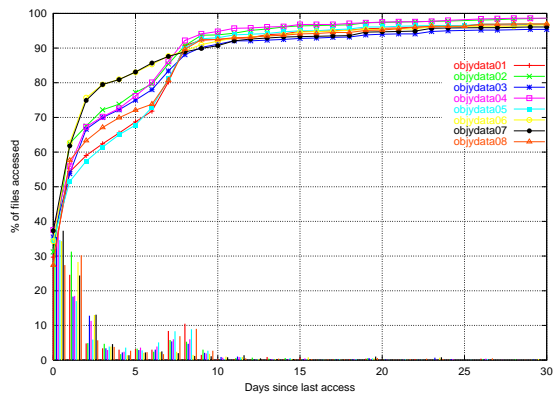

Figure 2: Access patterns on several hosts. Courtesy of Bill Weeks

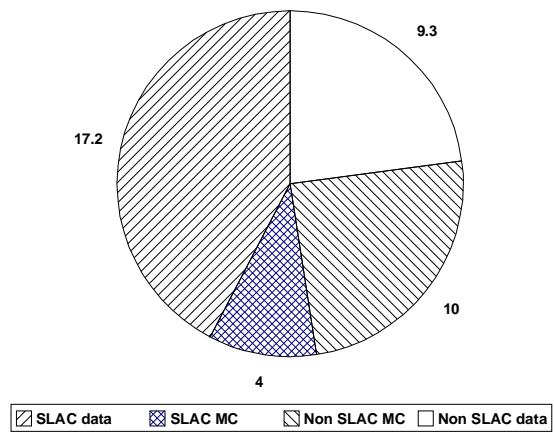

Figure 3: Run III data production (in TB since Nov'02).

## 5. IMPORT FROM REMOTE SITES

Since the beginning of the BaBar experiment, data produced at remote sites has been an important contribution to the total dataset. Starting from just a few external sites producing Monte Carlo data in 1999, now nearly 30 sites produce more than three quarters of all simulation data and one site (INFN-Padova) has set up remote data reprocessing farms. The ratios of current data volumes stored at SLAC are shown in Figure 3.

Four hosts at SLAC are dedicated to the data distribution activity, of which 2 hosts are used for simulation production import and 2 hosts are used for reprocessing data import. The import procedure uses an "active client" or "push" method, where remote sites initiate transfers to SLAC without handshaking or acknowledgment. SLAC is expected to be always available and ready to handle the incoming data. This allows a decoupling of server and client applications. The data transfer uses a special protocol, which defines where data is placed, format of metadata and messaging about transfer status. The message exchange is achieved via semaphore files, indicating the status of a data set: transfer in progress, transfer done, import in progress, import done.

Each import data set, or "import", usually corresponds to a production cycle, i.e. a set of runs allocated to a remote site to be processed. The total size of an import can be larger than the available disk space, and therefore, an import is usually split into many pieces which are handled individually. This also allows production at remote site to carry on while exporting closed databases and deleting them from local disk.

The SLAC import application detects new imports automatically and handles them. During an import, databases are attached to an import federation with integrity checking using an Objectivity/DB utility, then automatically migrated to HPSS and purged off disk. After importing, data is swept to the analysis federations. The import federation serves as a buffer before the analysis federation, where database integrity checks are performed. The import federations relieve the analysis ones from such relatively intrusive procedure. All imports are registered in a relational database with a web interface, where one can easily query all aspects of an import, such as the status of each import, size, number of events and runs, and also search for a database in case of problems. A command line interface also exists.

The same import handling application is used for both simulation and reprocessed data. At the same time, export applications are different. Export applications are usually set up to compress the data for higher throughput network transfer and calculate checksums for reliability.

## 6. DATA MANAGEMENT TOOLS

To efficiently work with BaBar's large data set, data management tools were developed. Due to the rapidly evolving event store technology and changing requirements, the process of development is almost continuous. One of the biggest tasks of BaBar database administration is the moving of data. This includes: sweeps from the production to analysis area, import of data coming from external sites, relocation of data within the analysis area, staging data in from tape, migrating data to tape, removing data from disk.

### 6.1. File Management Tools

File management tools are designed to meet the administrator's needs and consist of modules that perform the following tasks:

- Selection of databases
- Creation of command files to perform a task
- Execution of command files

Splitting tasks into these three stages allows the administrator to monitor each stage for possible errors, repeat each stage or defer running of some operation, when necessary. The selection module builds up a list of databases according to selected criteria including: database names, dbid, host, domain, authorization level, component type, modification time, readonly flag. Additionally, in reconstruction production, database can be selected by it's status: empty, in use, closed. The selection module supports include/exclude selection for database attributes and before/after selection on database time attributes.

Command files are created for each specified operation. This approach dictates, that each specified operation is performed on all databases before the next operation is performed. We found this approach convenient, as it allows greater flexibility in administration, and better protection from mistakes and failures. Furthermore, some operations require blocking user jobs from accessing the federations. Our approach allows us to reduce downtime to only those operations that do require the outage.

Typically, command files are run from the same application that produced them. For many operations, the command execution module is used to optimize the operation. Some operations, like staging, need to be executed on the hosts where the database resides, and the execution module automatically performs those operations on the required hosts. Some operations can be run in parallel, and the execution module allows for this. The number of parallel streams can be predefined for each operation, jobs can be run on different hosts in parallel too. The Execution module performs accounting and reporting of success and failures.

### 6.2. Sweep and Import Tools

Sweep tools are integrated into the file management package. Collection loading applications are part of the core BaBar database event store code and written in C++. The import tools for handling the data shipped from a remote site are packaged together with the tools that record imports. All tools are written Perl and shell script.

## 6.3. Maintenance Tools

With over 100 federations, automation of administration tasks is essential. Daily maintenance tasks include database schema upgrade and conditions snapshot transfers. For weekly maintenance, the following tasks are added: new user authorization, data and metadata backup. Before data backup is performed, the federation is inhibited to provide quiescence and so consistency of the data set on tape. Emergency maintenance tasks include - stop and start of the AMS, lock server and other services. Special tools were developed to deal with a number of federations in one command.

## 6.4. Monitoring Tools

### 6.4.1. Lock Monitoring

When an application crashes for any reason, it usually leaves an uncommitted transaction, which has to be rolled back in order to restore the federation to consistent state. A special Objectivity/DB utility "cleans up" such transactions and insures data consistency. In the analysis environment, a special application periodically runs and checks for "dead" transactions and cleans them up.

### 6.4.2. Hardware and Services Monitoring

The BaBar database infrastructure consist of over 100 servers, including data servers, lock servers, catalog servers. A monitoring job checks the availability of a server and reports to an administrator when a problem is detected. A number of services, including the AMS can be automatically restarted in case they fail.

## 7. SUMMARY

The BaBar experiment has managed to store over 700TB of data in Objectivity/DB databases. Maintaining maximum uptime and performance of such a large object oriented database system presents interesting challanges to the database administrators. With the tools and techniques that we have described we have managed to meet these challanges and provide and maintain a system that has been the basis for many ground-breaking and competitive physics analyses.

## Acknowledgments

The authors wish to thank the following people who contributed to BaBar Data Management. Jean-Noel Albert, Cristina Bulfon, Lawrence Mount, Hammad Saleem, Andy Salnikov, Yemi Adesanya, Andy Hanushevsky, Bill Weeks.